\documentclass[10pt,twocolumn,twoside,letterpaper]{IEEEtran}

\usepackage{geometry}
\usepackage{url}

\usepackage{amsmath}

\usepackage{multirow}

\usepackage{graphicx}
\usepackage{xcolor}
\newcommand{\rev}{}

\usepackage{cite}

\usepackage{comment}

\usepackage[modulo,switch]{lineno}

\begin{document}
\title{Front-end Electronics and Optimal Ganging Schemes for Single Photon Detection with Large Arrays of SiPMs in Liquid Argon}
\author{P.~Carniti, E.~Cristaldo, A.~Falcone, C.~Gotti, G.~Pessina, F.~Terranova
\thanks{The authors are with INFN Milano-Bicocca and University of Milano-Bicocca,
Piazza della Scienza 3, Milano, 20126 Italy.\\
e-mail: claudio.gotti@mib.infn.it}
}
\maketitle


\begin{abstract}
The operation of large arrays of silicon photomultipliers (SiPM) in tanks of noble liquids requires low noise, low power front-end amplifiers, able to operate reliably in the cryogenic environment.
A suitable amplifier needs to be paired with a proper SiPM ganging scheme, meaning the series/parallel combination of SiPMs at its input.
This paper presents a simple model to estimate the ganging scheme that gives the best signal to noise ratio once the basic electrical characteristics of the SiPM and amplifier are known.
To prove the validity of the model, we used an amplifier based on discrete components, which achieves a white voltage noise in the 0.25-0.37~nV/$\surd$Hz range at liquid nitrogen temperature, while drawing 2-5~mW of power.
Combined with the optimal ganging scheme obtained with the model, the amplifier demonstrated excellent single photon sensitivity up to 96 6x6~mm$^2$ SiPMs (total area 34.6~cm$^2$, S/N~$\simeq$~8-11).
The measured results are in a good match with calculated values, predicting the possibility to achieve a clear separation of photoelectron peaks {\rev also with larger areas}.
\end{abstract}

\section{Introduction}

\IEEEPARstart{S}{everal} next generation experiments in  astroparticle physics need cryogenic photon detection systems to detect scintillation light produced by particle interactions in large time projection chambers filled with noble liquids, typically liquid argon (LAr) at 87~K or liquid xenon at 165~K.
Notable examples that use LAr are DarkSide \cite{DarkSide} and DUNE \cite{DUNE}.
Nowadays silicon photomultipliers (SiPM) are often the sensors of choice for this class of applications, since they offer single photon sensitivity, low operating voltage and good mechanical reliability in a compact footprint.
The much larger rate of thermally generated dark counts is their main drawback compared to vacuum-based photon detectors, but it is made negligible by operation in a cryogenic environment.
Another drawback is their larger capacitance per unit area, which needs to be taken into account when designing the front-end circuitry.
From a physics standpoint, given the nature of the events of interest, reading out each SiPM separately is not required.
On the contrary, optical techniques can be used to collect photons on a larger area ($\sim 0.1-1$~m$^2$) and guide them, with the highest possible efficiency, towards a smaller but still relatively large area (tens of cm$^2$) equipped with SiPMs, which constitutes a single channel.
An example of this approach is the ARAPUCA technology to be used in DUNE \cite{ARAPUCA1, ARAPUCA2}.

Given the size of the cryostats where these experiments operate, the front-end circuits are also placed in LAr, close to the photon detectors. They need to provide adequate signal to noise ratio (S/N), bandwidth and dynamic range, and drive the signals outside the cryostat through cables or optical fibers.
Digitization in cold might be an option, although it offers its own challenges.
The front-end circuits need to operate with a power density that does not cause bubbling in LAr.
As expected, the S/N performance of the readout chain is determined by the characteristics of the SiPM and of the first amplification stage, and how the former are connected to the latter.

\section{SiPM ganging}

It is in principle possible to segment the readout channels and equip each SiPM, or small group of SiPMs, with an amplifier, then sum the signals after this first amplification stage.
The noise of each sub-group would be uncorrelated from the others: as $N$ sub-groups are combined, the S/N scales as $1/\sqrt{N}$.
From a practical point of view this can be challenging, because the total power consumption of the amplifiers in a readout channel would scale as $N$, while each amplifier still needs to satisfy S/N and bandwidth requirements that generally conflict with low power operation.

Another option is to combine (gang) a large number of SiPMs and read them out with a single front-end amplifier.
The easiest way to do this is to connect all SiPMs in parallel.
Let us indicate by $Z_c$ the source impedance of a single SiPM cell.
It can be approximated by the junction capacitance of the SPAD $C_d$ in series with its quenching resistance $R_q$.
When a cell fires, it generates a current signal that decays exponentially with time constant $\tau_s = C_d R_q$. 
It is of course possible to consider more realistic models, which result in a better prediction of the signal shape, especially in their fast components \cite{SiPMmodel1,SiPMmodel2,SiPMmodel3,SiPMmodel4}.
{\rev Since the evaluation of the S/N is typically done on the integral of SiPM signals, their high frequency features have little or no effect on the S/N. Hence the basic SiPM model will be sufficient within the scope of this paper.}
If each SiPM has $M$ cells, and $N$ SiPMs are connected in parallel, the total source impedance presented to the readout circuit is $Z_a = Z_c/MN$.
This can be modeled as a capacitance $C_a=C_d MN$ in series with a resistance $R_a = R_q/MN$.
Note that the time constant defined by $R_a$ and $C_a$ is still $\tau_s$, same as a single SiPM cell.
Considering a photosensitive area of tens of cm$^2$, given the relatively high capacitance per unit area of SiPMs, of the order of 35-60~pF/mm$^2$ depending on technology, the total source capacitance $C_a$ of the array can range up to 100~nF, while the resistance $R_a$ becomes negligible.
It is then unrealistic to reach the conditions where the front-end capacitance matches the capacitance of the detector and optimal S/N is achieved \cite{OptimalSN}.
It is possible to work in mismatched conditions, where the source capacitance of the SiPM array is much higher than the input capacitance of the front-end amplifier, and noise is dominated by the series (voltage) noise of the latter.
{\rev As a consequence, the equivalent noise charge is directly proportional to the total capacitance of the SiPM array.} In other words, the S/N scales as $1/N$ with the number of SiPMs.
Clearly, an amplifier with exceptionally low series noise is required.
This is the case of the DUNE FD1 photon detection system, where SiPMs are connected in parallel for a total surface of 17~cm$^2$ per readout channel \cite{ColdAmpDUNE1, ColdAmpDUNE2}.
An amplifier with similar noise characteristics has been developed for the DarkSide experiment \cite{DarkSideAmp1, DarkSideAmp2}.

\begin{figure}[t]
\centering
\includegraphics[width=\linewidth]{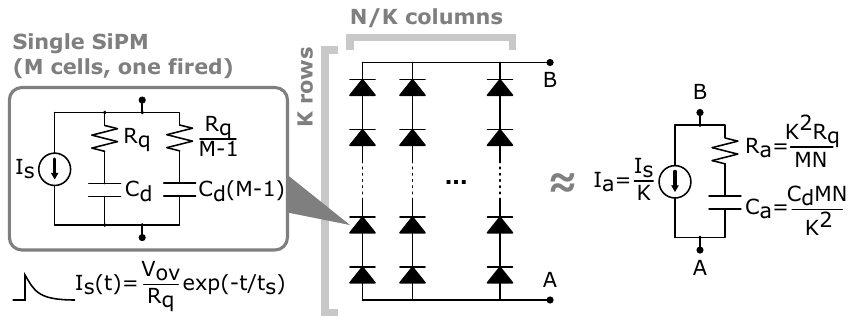}
\caption{Simplified model for $N$ ganged SiPMs, arranged as the parallel combination of $N/K$ sub-groups, each of $K$ SiPMs in series.
The fact that the signal is attenuated by a factor $K$ without changing shape is true assuming that the array is read out by an amplifier with negligible input impedance.}
\label{fig:GangEquivalent}
\end{figure}

A way to mitigate the impedance mismatch and reduce the weight of the series noise is to connect groups of SiPMs in series, as shown in Fig. \ref{fig:GangEquivalent}.
Let us still denote with $N$ the total number of SiPMs in the array.
If $K$ SiPMs are connected in series to form a group, and then $N/K$ identical groups are connected in parallel (requiring $N/K$ to be an integer), the source impedance of the array of $N$ SiPMs becomes 
\begin{equation}
Z_a (s) = \frac{K^2 Z_c}{MN} = \frac{K^2 R_q}{MN} \frac{1+s\tau_s}{s \tau_s}
\label{eq:Za}
\end{equation}
which is a factor $K^2$ higher than the case where the same number of SiPMs are all connected in parallel.
The impedance of the array can also be expressed as
\begin{align}
\nonumber Z_a(s) = R_a+\frac{1}{s C_a}=R_a \frac{1+s\tau_s}{s \tau_s}\\
R_a = \frac{K^2 R_q}{MN}
\hspace {0.3cm}
C_a = \frac{C_d MN}{K^2}
\label{eq:Za2}
\end{align}
This arrangement effectively increases the impedance that the SiPM array presents to the front-end amplifier at signal frequencies, compared with the previous case with all SiPMs in parallel.
Eq. (\ref{eq:Za}) and (\ref{eq:Za2}) remain valid also when the basic element of the array (the single SiPM of Fig. \ref{fig:GangEquivalent}) is replaced by the parallel combination of a few SiPMs, as long as the number of rows in series is $K$ and the total number of SiPMs is $N$.
This scheme was pioneered in the framework of the MEG II experiment \cite{GangMEGII}, although with a small SiPM area and without a cold amplification stage, and in the DarkSide and nEXO experiments  \cite{GangDarkSide,GangnEXO}, with 6 cm$^2$ of SiPM area read out by a single amplifier.
{\rev In practice, the connection of SiPMs in series generally requires adding resistors in parallel with the SiPMs, to ensure that the bias voltage is equally distributed. Also, the total bias voltage that needs to be provided is $K$ times larger than in a parallel connection.}

A SiPM is a two-terminal device.
As shown in Fig. \ref{fig:GangEquivalent}, each cell can be represented by its Norton equivalent circuit: an impedance $Z_c(s) \simeq R_q + 1/sC_d$ in parallel with a current generator $I_s$.
{\rev A single SiPM cell that fires at $t=0$ gives a signal}
\begin{equation}
I_s(t) = \frac{Q}{C_d}  \exp (-t/\tau_s) = \frac{V_{ov}}{R_q} \exp (-t/\tau_s)
\label{eq:IsTime}
\end{equation}
where $V_{ov}$ is the overvoltage and $\tau_s = R_q C_d$ is the fall time constant of the signals.
Again, we are using a simplified SiPM model, but it is enough for the purpose.
Assuming that the SiPM array is read out by a transimpedance amplifier with negligible input impedance, the nodes labelled B and A in Fig. \ref{fig:GangEquivalent} are held at constant voltage.
The single photoelectron signal current $I_s$ is loaded by one SiPM in parallel with the series of the other $K-1$ SiPMs. Hence it generates a voltage across the SiPM it originates from:
\begin{equation}
V_s= I_s \frac{Z_s (K-1) Z_s}{Z_s + (K-1) Z_s} = I_s Z_s \frac{K-1}{K}
\end{equation}
A part of the signal current will flow through the other $K-1$ SiPMs, and will be collected by the amplifier:
\begin{equation}
I_a = \frac{V_s}{(K-1) Z_s} = \frac{I_s}{K}
\label{eq:Ia}
\end{equation}
which, considering (\ref{eq:IsTime}), in the Laplace domain becomes
\begin{equation}
I_a(s) = \frac{V_{ov}}{K R_q} \frac{\tau_s}{1+s \tau_s}
\label{eq:IaLaplace}
\end{equation}
In essence, as long as the SiPM array is read out by an amplifier with negligible input impedance, the single photoelectron signal does not change shape due to the series ganging configuration, but is attenuated by a factor $K$.
(The configuration with all SiPMs in parallel can be seen as the special case where $K=1$. Eq. (\ref{eq:Za}) to (\ref{eq:IaLaplace}) are still valid, $V_s$ vanishes and the entire SPAD current $I_s$ reaches the amplifier.)

A variant of this approach is to connect the SiPMs through resistors and capacitors, so that their bias voltage (DC) can be distributed in parallel to all SiPMs, while the signal (AC) sees groups of SiPMs in series.
This scheme has been used for the measurements presented later in this paper, so more details will follow.
From the point of view of the S/N, this configuration is equivalent to the series ganging discussed above, while resulting more practical for SiPM biasing.
The impedance of the SiPM array at signal frequencies is still given by (\ref{eq:Za}) and (\ref{eq:Za2}), and the signal current is still given by (\ref{eq:IaLaplace}).
This approach has been called hybrid ganging \cite{GangHybrid}, again pioneered with a few SiPMs (1.5~cm$^2$) in the context of MEG II, and then adopted on a larger scale (28.8~cm$^2$ per readout channel) for the second far detector of DUNE \cite{GangDUNEVD}.

These ganging schemes rely on the fact that all SiPMs in each array have very similar values of breakdown voltage, quenching resistance and SPAD capacitance.
This is a reasonable requirement, at least from SiPMs that come from the same production batch, and sorting is not typically required.

\section{Single photoelectron resolution}
\label{sec:SinglePeResolution}

The capability of resolving single photoelectrons is quantified by the S/N.
We can define the S/N at the output of the entire readout chain (including analogue or digital filtering that occurs after the amplifier) as the amplitude of single photoelectron signals divided by the RMS baseline noise.
If all SiPMs are identical, and neglecting possible effects due to afterpulses, all photoelectron peaks in the spectrum are expected to have the same width, which is determined by the intrinsic noise sources of SiPM and amplifier, to be detailed later.
In other words, in the SiPM charge spectra obtained in low light conditions, the S/N will be the distance between peaks divided by their sigma.
If there is a spread in SiPM characteristics across the array, then the peaks at one and more photoelectrons might be larger than the baseline peak at zero photoelectrons.
We assume that this effect can be made negligible by matching the SiPMs in the array, and we do not include it in the definition of the S/N, which will only consider electronic noise sources.

\begin{figure}[t]
\centering
\includegraphics[width=\linewidth]{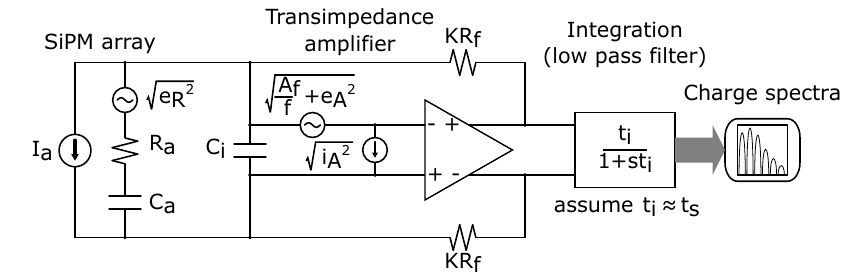}
\caption{Small signal model of the signal readout chain, comprised of a transimpedance amplifier and a low pass filter that models signal integration. Noise sources are included. SiPM biasing is not shown.}
\label{fig:ReadoutModel}
\end{figure}

The signal readout chain can be modeled as shown in Fig. \ref{fig:ReadoutModel}.
The SiPMs are assumed to be biased at overvoltage $V_{ov}$ above breakdown, but the biasing circuit is not considered in the scheme.
The signal current $I_a$ is read out by a transimpedance amplifier and converted to a voltage.
{\rev A fully differential readout scheme is assumed, but the results will not depend on this choice. A single-ended readout would give the same S/N, provided that all other quantities remain the same.}
 The differential gain is $2 K R_f$.
Starting from a reference value $R_f$ chosen to get the proper gain with all SiPMs in parallel ($K=1$), we assume that the value of the feedback resistor, hence the gain, is scaled proportionally to $K$ to compensate the signal loss with $K$, as expressed by (\ref{eq:Ia}).
The output of the amplifier is integrated to recover the charge in each signal, and the results populate the spectra.
Integration over infinite time could be represented as a low pass filter with a single pole at zero frequency, but cannot clearly be implemented in practice.
Integration on a time window of finite length $T$ can be represented as a single pole low pass filter with cutoff frequency $f_i \simeq 0.443 / T$, or $\tau_i \simeq T/2.78$ {\rev (proof given in Appendix A)}.
The value of the integration window $T$ should be equal to a few SiPM time constants $\tau_s$, so that all the relevant signal is integrated.
As a reasonable approximation that greatly simplifies calculations, we will set here the integration window to $T \simeq 3 \tau_s$, so that $\tau_i \simeq \tau_s$.
Using (\ref{eq:IaLaplace}), the single photoelectron signal at the output of the readout chain is then\footnote{$S(t)$ is a voltage integrated in time and has dimensions of V$\cdot$s. The same happens for the noise, and the S/N will be a dimensionless quantity.}
\begin{align}
\nonumber S(s) &= 2 K R_f I_a(s) \frac{\tau_i}{1+s\tau_i} \\
&\simeq  2 R_f \frac{V_{ov}}{R_q} \left( \frac{\tau_s}{1+s\tau_s} \right)^2
\end{align}
Converting back to time domain, we obtain the single photoelectron signal at the output of the readout chain. The maximum of the filtered output signal corresponds to the instant when the integration window is optimally centered on the signal ($t = \tau_s$):
\begin{equation}
S_{MAX} =  2 R_f \frac{V_{ov}}{R_q} \frac{\tau_s}{\exp(1)}
 \label{eq:SMAX}
\end{equation}
This quantity is the distance between peaks in the charge spectrum, and goes at the numerator in the S/N.

The noise sources shown in Fig. \ref{fig:ReadoutModel} are the voltage noise of the amplifier, composed of a white part $e_A^2$ and a flicker component $A_f/f$, the current noise of the amplifier $i_A^2$, assumed white, and the thermal (white) noise of the resistance of the SiPM array, $e_R^2 = 4 k_b T R_a$.
Capacitor $C_i$ is neglected here, and will be discussed in section \ref{sec:MeasurementSetup}.
As long as $R_f$ is much larger than the source impedance $Z_a$, the voltage noise sources $e_A^2$ and $e_R^2$ have the same transfer function:
\begin{align}
\nonumber
N_{w}^2(s) &=  \left|\frac{2 K R_f}{Z_a (s)}
\frac{\tau_i}{1+s\tau_i} \right|^2  \left(e_A^2 + e_R^2 \right) \\
& \simeq   \left| \frac{2 K R_f}{R_a}
\frac{s \tau_s^2}{\left(1+s\tau_s \right)^2} \right|^2  \left(e_A^2 + e_R^2 \right)
\label{eq:Nw}
\end{align}
where (\ref{eq:Za2}) was used for $Z_a(s)$.
The flicker noise needs to be treated separately and gives
\begin{equation}
N_{1/f}^2(s) \simeq \left| \frac{2 K R_f}{R_a}
\frac{s\tau_s^2}{\left(1+s\tau_s \right)^2} \right|^2   \frac{2 \pi A_f}{\left| s \right|}
\label{eq:N1f}
\end{equation}
while the current noise has a different transfer function, the same as the SiPM signal:
\begin{align}
N_{i}^2(s) &=  \left|{2 K R_f} \frac{\tau_i}{1+s\tau_i} \right|^2 i_A^2 \simeq
\left| {2 K R_f} \frac{\tau_s}{1+s\tau_s} \right|^2 i_A^2
\label{eq:Ni}
\end{align}
From the above expressions, by Parseval's theorem, {\rev and using known integrals given in Appendix B}, we can calculate the RMS fluctuation.
The result, after all sources are summed in quadrature, is
\begin{equation}
N_{RMS} =  \frac{2 K R_f}{R_a} \left[
\frac{\tau_s}{8}(e_A^2 + e_R^2) + \frac{\tau_s^2}{2}A_f + \frac{\tau_s}{4} R_a^2 i_A^2
 \right]^{\frac{1}{2}}
 \label{eq:NRMS}
\end{equation}
This is the quantity that goes at the denominator in the S/N.

Dividing (\ref{eq:SMAX}) by (\ref{eq:NRMS}), and using $R_a$ from (\ref{eq:Za2}), the S/N can be written as
\begin{equation}
S/N= \frac{V_{ov} \sqrt{8 \tau_s}}{\exp(1)}
\frac{K}{MN} \frac{1}{\sqrt{e_A^2 + e_R^2 + 4 \tau_s A_f + 2 R_a^2 i_A^2}} 
\label{eq:SN}
\end{equation}
{\rev The S/N in (\ref{eq:SN}) quantifies the capability of resolving single photoelectron signals from the baseline, when the latter fluctuates because of intrinsic electronic noise sources. It does not include interference from the environment (which depends on how well the sensitive parts of the system are shielded and isolated from a specific environment) nor the effect of dark counts of the SiPM array (which may result in pile-up of single photoelectron signals, but is typically negligible at cryogenic temperature).}

To contribute to the total RMS noise in (\ref{eq:NRMS}), the corner frequency of the amplifier $A_f/e_A^2$ should be comparable with $1/2 \pi \tau_s$.
Given a typical value of the SiPM time constant of $\sim$100~ns, this implies a corner frequency in the 100~kHz range for the flicker noise to be relevant.
As for the current noise contribution, its weight depends on the value of $R_a$, which for a given SiPM depends on the ganging configuration.
{\rev Even for relatively high values of $K$, we expect $R_a$ to be at most in the $100\ \Omega$ range, which likely makes the current noise negligible compared with the voltage noise sources. We will proceed assuming it is negligible, but this will need to be verified for each particular case.}
If we neglect the current noise, and use the explicit form of $e_R^2 = 4 k_b T R_a$, Eq. (\ref{eq:SN}) becomes
\begin{equation}
S/N= \left[\frac{V_{ov} \sqrt{8 \tau_s}}{\exp(1)}\right]
\left[\frac{K}{M N} \frac{1}{\sqrt{e_A^2 + 4 k_b T \frac{K^2 R_q}{M N} + 4 \tau_s A_f}} \right]
\label{eq:SN2}
\end{equation}
The first factor in (\ref{eq:SN2}) does not depend on the ganging configuration, and is not particularly interesting in this context.
Its dependence on $V_{ov}$ is trivial: higher overvoltage results in higher single photoelectron signal without affecting the baseline noise, although a higher $V_{ov}$ results in higher probability of crosstalk and afterpulses, so it cannot be a handle to increase the S/N ad libitum.
The dependence on $\tau_s$ seems to suggest that longer $\tau_s$ (and therefore longer $\tau_i$) will improve the S/N, as is to be expected in the presence of series white noise. But this, besides being impractical, does not affect the $1/f$ noise.
Longer integration times will also affect the weight of afterpulses on the S/N and might degrade the timing resolution.
The second factor in (\ref{eq:SN2}) depends on the ganging configuration, and leads to interesting considerations.
Albeit obtained with many simplifications, this expression is meaningful because it highlights two S/N regimes.
{\rev The boundary between the two regimes corresponds to the value of $K$ for which the first and last term at the denominator of the second factor in (\ref{eq:SN2}), which do not depend on $K$, are equal to the second term, which depends on $K$.}
Let us define
\begin{equation}
\tilde{K} = \sqrt{\frac{{\left(e_A^2 + 4 \tau_s A_f\right) MN}}{{4 k_b T R_q}}}
\label{eq:tildeK}
\end{equation}
{\rev For $K<<\tilde{K}$ Eq. (\ref{eq:SN2}) becomes}
\begin{equation}
S/N \simeq \frac{V_{ov} \sqrt{8 \tau_s}}{\exp(1)}
\frac{K}{M N \sqrt{e_A^2 + 4 \tau_s A_f}}
\label{eq:SNsmallK}
\end{equation}
The dominant noise contributor is the voltage noise of the amplifier, while the thermal noise of the SiPM quenching resistors is negligible.
For large $N$ the S/N scales as $1/N$.
In other words, the S/N is inversely proportional to the total input capacitance, as was discussed for the case of parallel ganging.
The S/N is also proportional to $K$, signaling that, as long as $K<\tilde{K}$, connecting more SiPMs in series will improve the S/N.
{\rev On the other hand, for $K>>\tilde{K}$ Eq. (\ref{eq:SN2}) becomes}
\begin{equation}
S/N \simeq \frac{V_{ov} \sqrt{8 \tau_s}}{\exp(1)}
\frac{1}{\sqrt{M N 4 k_b T R_q}}
\label{eq:SNlargeK}
\end{equation}
Now the noise of the amplifier is negligible compared with the thermal noise of the equivalent resistance of the SiPM array.
The latter is still a voltage noise source, but as $N$ increases the S/N scales as 
$1/\sqrt{N}$.
This does not mean that the noise of the amplifier does not play a role: since $\tilde{K}$ depends on  $e_A^2$ and $A_f$, the value of $K$ required to reach this regime can become impractically high if the amplifier noise is not low enough.
Interstingly, once $K$ is large enough to reach this regime, the S/N expressed by (\ref{eq:SNlargeK}) does not depend on $K$ anymore, and there is no point in increasing $K$ further.
On the contrary, a too large $K$ can have detrimental effects: the stronger signal attenuation, although compensated by the larger gain of the transimpedance amplifier, can make the readout circuit more sensitive to environmental interference. Also a large $K$ can lead to the point where the current noise of the amplifier cannot be neglected anymore, since its contribution is proportional to $R_a$.

In conclusion, we may state that once the basic characteristics of the SiPM array and of the amplifier are fixed and known, the optimal ganging configuration, or in other words the value for $K$ that gives the best S/N, corresponds to the first integer value larger than $\tilde{K}$ as given by (\ref{eq:tildeK}), with the additional constraint that $N/K$ must be integer as well.
This was calculated under the assumption that the current noise of the amplifier could be neglected, which will be likely true in most cases.
{\rev If this were not the case, we would need to go back to (\ref{eq:SN}) to find the value of $K$ that maximizes the S/N.}

\section{Measurement setup}
\label{sec:MeasurementSetup}

Having discussed the model of the system on paper, we are going to put the theory to the test with measurements on specific instances of SiPMs, amplifiers and ganging configurations.

\subsection{SiPM model}

We used SiPMs developed by Fondazione Bruno Kessler (FBK) for the DUNE experiment, as a pre-production batch of devices for the FD1 photon detection system.
They are a variation of the NUV-HD-Cryo technology developed for the DarkSide experiment \cite{NUVHDCryo}, with 54~$\mu$m cell pitch and triple trenches between cells.
Each SiPM has a surface of 6$\times$6 mm$^2$ and $M=11188$ cells.
The nominal cell capacitance $C_d$ is close to 200~fF, the quenching resistor $R_q$ is just above 3~M$\Omega$ at cryogenic temperature, for a total resistance per SiPM of about 300 $\Omega$. The recovery time constant $\tau_s$ is about 600~ns, relatively long compared to other SiPM types.
In the measurements presented here, the signal fall time constant was shortened to 300~ns by an AC coupling described in subsection \ref{subsec:pulsedLED}.
The tests were performed in liquid nitrogen (LN2) at 77~K, close to the temperature of LAr.
The SiPM breakdown voltage in LN2 is 27.1~V, and they are typically operated at $V_{ov}=4.5$~V overvoltage, which results in a detection efficiency of 45\%.

\subsection{Amplifier variants}

\begin{figure}[t]
\centering
\includegraphics[width=\linewidth]{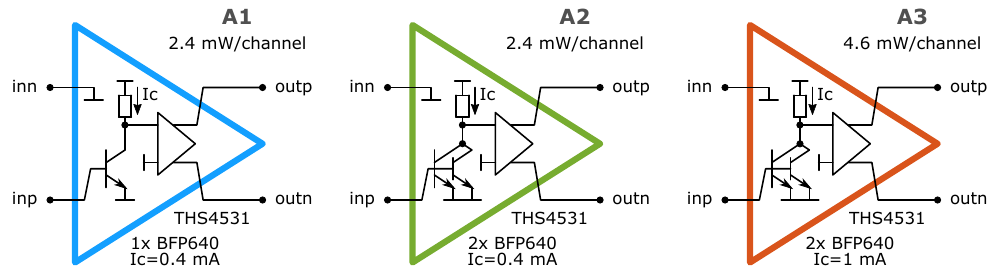}
\includegraphics[width=0.8\linewidth]{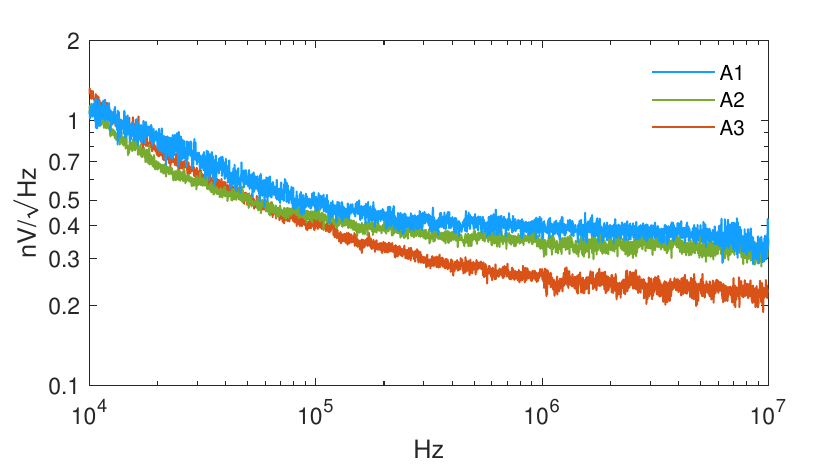}
\caption{Simplified schematic of the three amplifier variants and their input-referred voltage noise spectra measured at 77~K.}
\label{fig:AmpNoise}
\end{figure}

We used three amplifiers, all based on a design developed for DUNE FD1 \cite{ColdAmpDUNE2}.
The input device is a BFP640 SiGe heterojunction bipolar transistor, which gives low series noise at low collector currents, followed by a THS4531 fully differential opamp, which provides the bulk of the loop gain and drives the differential output lines to the warm second stage.
Fig. \ref{fig:AmpNoise} presents the three amplifier variants with their voltage noise spectra.
The baseline version (A1) uses a single SiGe transistor operated at \mbox{0.4 mA} collector current. The white voltage noise in LN2 is \mbox{0.37 nV/$\surd$Hz}, and the base current is about \mbox{1 $\mu$A}, giving a current noise of about \mbox{0.6 pA/$\surd$Hz}.
The entire amplifier draws a total current of \mbox{0.7 mA} from a single \mbox{3.3 V} supply, which results in a power consumption of 2.4~mW.
As shown in a previous work \cite{ColdAmpDUNE1}, the noise at cryogenic temperature cannot be completely accounted for by the transconductance term. An additional contribution is present, and was attributed to the base spreading resistance of the transistor $R_{bb'}$, with a value of about \mbox{20 $\Omega$}:
\begin{equation}
e_A^2 = \frac{2 k_b T}{g_m} + 4k_b T R_{bb'} =
4 k_b T \left[ \frac{k_b T}{2 q I_c} + R_{bb'} \right]
\end{equation}
Increasing the collector current above \mbox{0.4 mA} does not lead to a significant noise reduction, indicating that at the noise at \mbox{0.4 mA} is already limited by the $R_{bb'}$ term.
Instead of increasing the current, it is more effective to connect two transistors in parallel (A2).
Even at the same total collector current of \mbox{0.4 mA}, or \mbox{0.2 mA} in each transistor, the effective value of $R_{bb'}$ is halved, which leads to a lower overall $e_A$ of \mbox{0.33 nV/$\surd$Hz} at the same power consumption of 2.4~mW.
The total base current is still about \mbox{1 $\mu$A} and the current noise about \mbox{0.6 pA/$\surd$Hz}.
The third amplifier variant A3 still has two transistors in parallel, with a total collector current increased to 1 mA.
This brings the power consumption to 4.6~mA, with a further decrease in noise, which goes down to \mbox{0.25 nV/$\surd$Hz}.
The total base current is now a few \mbox{$\mu$A} and the current noise is close to \mbox{1 pA/$\surd$Hz}.
Fig. \ref{fig:AmpNoise} shows the input-referred voltage noise spectra for the three amplifier variants between 10~kHz and 10~MHz.
Note that due to the very low white component, variant A3 has a higher corner frequency, just above 100~kHz, which combined with the relatively long $\tau_s$ of the SiPMs (see previous subsection) makes the flicker component not completely negligible.
The $1/f$ coefficient $A_f$ can be estimated to be $10^{-14}$~V$^2$.
Direct evaluation with (\ref{eq:NRMS}) shows that the contribution is anyway subleading compared to the white voltage noise components.

\subsection{Ganging configurations}

\begin{figure}[t]
\centering
\includegraphics[width=0.8\linewidth]{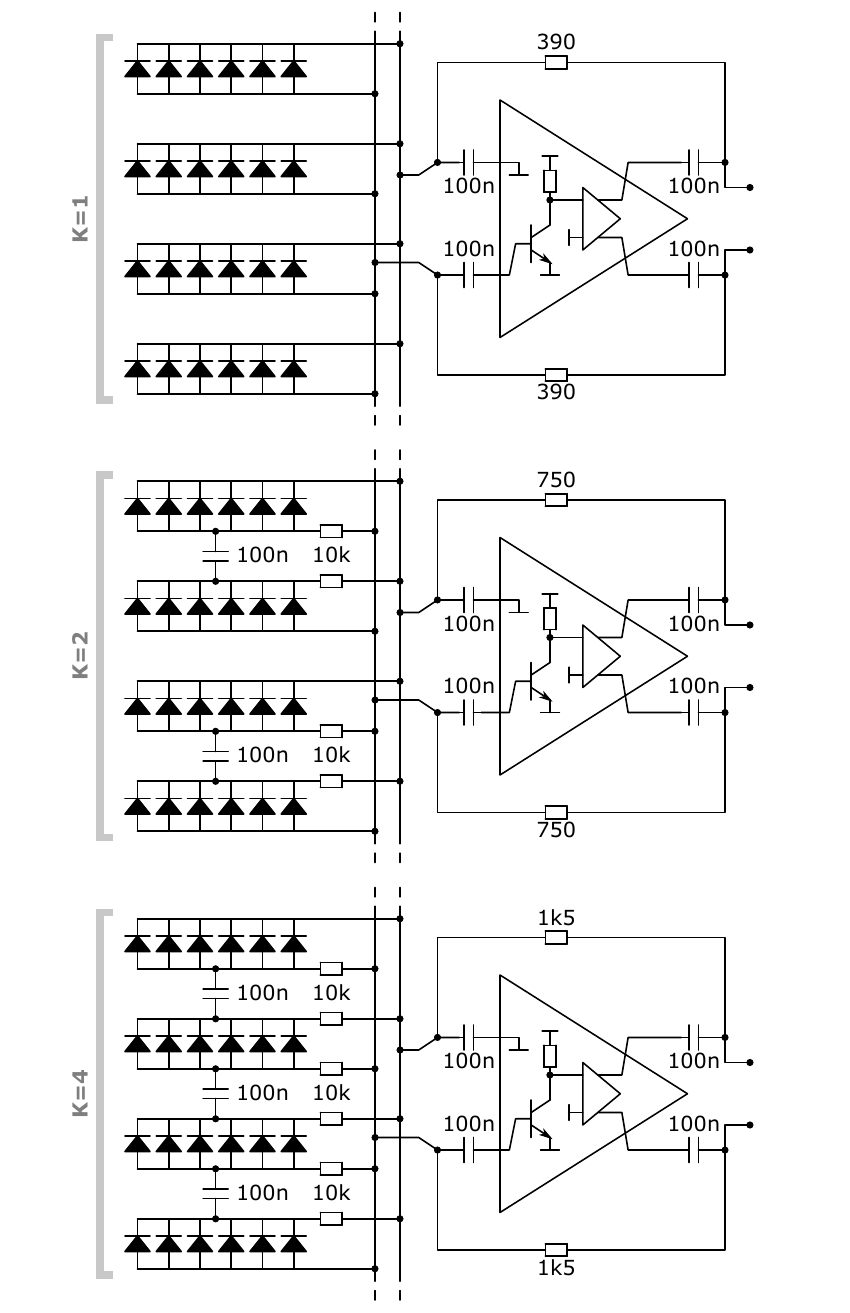}
\caption{Hybrid ganging with $K=1$, $K=2$, $K=4$ among subgroups of 6 SiPMs in parallel. All schemes are shown for 24 SiPMs, but have been extended to 48, 72 and 96 by connecting in parallel groups of 24. The bias for the SiPMs is applied in DC to the same pair of wires used in AC for signal amplification and transmission. Note that the feedback resistor scales with $K$ to increase the gain and compensate the attenuation due to the series connection of SiPMs.}
\label{fig:GangingSchemesWithAmp}
\end{figure}

Groups of 6 SiPMs came already mounted on PCBs with a common cathode pin, which means that SiPMs were necessarily connected in parallel in subgroups of 6.
Those subgroups were then connected in three different hybrid ganging configurations, shown in Fig. \ref{fig:GangingSchemesWithAmp}.
This scheme was applied with $K=1$ (all SiPMs in parallel), $K=2$ and $K=4$.
For $K>1$, all SiPMs are still in parallel in DC through 10~k$\Omega$ resistors, so they are biased at the same voltage.
The subgroups are then connected in AC through 100~nF capacitors\footnote{\rev We used C0G ceramic capacitors rated for 100~V in 1206 package.}, so that at signal frequencies they are effectively connected in series.
For each ganging scheme Fig. \ref{fig:GangingSchemesWithAmp} shows 24 SiPMs.
The scheme can be extended to 48, 72, 96 by connecting in parallel groups of 24.
The amplifier is coupled in AC at the inputs and outputs, to allow to use the same wire pair for SiPM biasing and signal transmission.
The capacitors are inside the feedback loop of the amplifier, and their value is large enough that it does not affect the signal shape in a significant way.

Knowing the relevant quantities for SiPMs and amplifiers, we can use (\ref{eq:tildeK}) to calculate the value for $\tilde{K}$ for different corner cases.
Using A1 we get $\tilde{K} = 1.6$ for 24 SiPMs and $\tilde{K} = 3.3$ for 96 SiPMs.
Using A3 we get $\tilde{K} = 1.2$ for 24 SiPMs and $\tilde{K} = 2.3$ for 96 SiPMs.
With $K$ ranging up to 4, all amplifier variants reach the ``optimal'' regime as discussed in section \ref{sec:SinglePeResolution}.
At this point we can also verify the assumption that the current noise of the amplifier can be neglected.
From (\ref{eq:NRMS}), its weight becomes relevant when $R_a > 1$~k$\Omega$.
Considering the corner with lowest $N=24$, this would require $K>8$, which is far above all ganging schemes considered here.
The margin becomes larger as $N$ is increased.

\subsection{Pulsed LED and waveform acquisition}
\label{subsec:pulsedLED}

The differential outputs of the amplifier were converted to single-ended by a second stage at room temperature, further amplified by a factor 10 and acquired by the oscilloscope.
The conversion to single-ended was done by a transformer (H1164NL) as discussed in a previous work \cite{ColdAmpDUNE2}. As a side effect, this introduced a zero (equivalent to an AC coupling) that shortened the effective fall time constant of the signals from 600~ns to about 300~ns.
A pulsed LED illuminated the SiPM array through an optical fiber. The intensity of the LED was set so that the array received just a few photoelectrons in average for each LED pulse.
The oscilloscope was triggered by the LED driver.
The waveforms were sampled at 2~ns steps for 2~$\mu$s (1000 points per waveform).
Each dataset consisted of 1000 waveforms.
The single photoelectron spectra were built by integrating the signals over a 1~$\mu$s window ($\tau_i \simeq 300$~ns) and populating histograms with the resulting values.

\begin{figure}[t]
\centering
\includegraphics[width=0.8\linewidth]{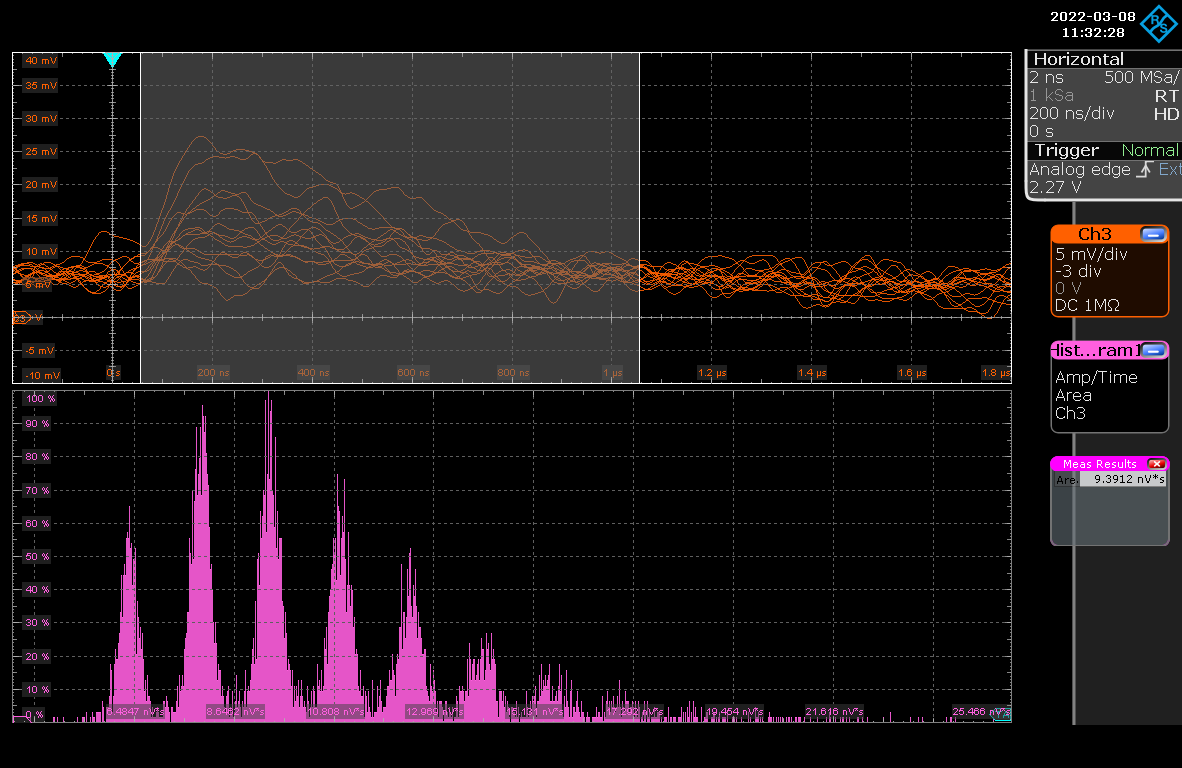}
\caption{Signals corresponding to a few photoelectrons as acquired by the oscilloscope. The histogram is calculated by integrating over the 1~$\mu$s window highlighted. This measurement was obtained with the amplifier variant A2 and 96 SiPMs in hybrid ganging with $K=2$, and gave S/N~$=8$.}
\label{fig:OscilloScreenshot}
\end{figure}

Fig. \ref{fig:OscilloScreenshot} shows an example of oscilloscope acquisition, including the histogram of waveform integrals on a 1~$\mu$s window, where peaks corresponding to 0, 1, 2... photoelectrons are clearly separated.
The waveforms of each dataset were saved and analyzed offline in a similar way, and the gaussian peaks were fitted to determine their mean $\mu_i$ and standard deviation $\sigma_i$, with $i=0,\ 1,\ 2$.
The S/N was calculated as $(\mu_1-\mu_0)/\sigma_0$.
In the case of the dataset of Fig. \ref{fig:OscilloScreenshot} this gave S/N~$=8$.

\subsection{Correction to S/N due to a 10~nF input capacitor}
\label{subsec:10nFcap}

In all the measurements presented here, a 10~nF capacitor in series with a 1~$\Omega$ resistor were connected across the inputs of the amplifier.
This was done for the practical necessity of maintaining the stability of the feedback loop independently of the number of SiPMs in the array.
If the feedback loop was optimized for a fixed given number of SiPMs these components could be omitted.
Fig. \ref{fig:ReadoutModel} included such capacitor $C_i$, which was neglected in the general discussion of section \ref{sec:SinglePeResolution}, but has an impact on the S/N predictions.
We can continue neglecting the small 1~$\Omega$ resistor and focus on the effect of the capacitor.
The presence of $C_i$ affects the amplifier noise sources $e_A^2$ and $A_f/f$, which see a different noise gain, but has no effect on $e_R^2$ and $i_A^2$. Eq. (\ref{eq:Nw}) and (\ref{eq:N1f}) need to be modified accordingly.
The total impedance connected to the the input of the amplifier becomes
\begin{align}
\nonumber Z_i(s) &= Z_a(s) || \frac{1}{s C_i} \\
\nonumber & = Z_a(s) \left[ 1+ \left(R_a C_i/\tau_s\right) \left(1+ s \tau_s \right) \right]^{-1}\\
& \simeq Z_a(s) \left[ 1+ R_a C_i/\tau_s \right]^{-1}
\label{eq:Zialpha}
\end{align}
The approximation in (\ref{eq:Zialpha}) neglects a factor that caused the noise gain to diverge at high frequency. This is done for the sake of simplicity, and is justified under the assumption that other factors will limit the noise gain at high frequency, like the presence of the 1~$\Omega$ small resistor in series with $C_i$ and the finite closed loop bandwidth of the amplifier.
Replacing $Z_a(s)$ with $Z_i(s)$ in the amplifier noise terms, (\ref{eq:NRMS}) can be rewritten as
\begin{equation}
N_{RMS} \simeq \frac{2 K R_f}{R_a} \left[
\frac{\tau_s}{8}(\alpha^2 e_A^2 + e_R^2) + \frac{\tau_s^2}{2} \alpha^2 A_f + \frac{\tau_s}{4} R_a^2 i_A^2
 \right]^{\frac{1}{2}}
 \label{eq:NRMS2}
\end{equation}
where $\alpha =  1+ R_a C_i/\tau_s$ depends on the ganging configuration through $R_a$.
The correction affects measurements taken with higher values of $R_a$, hence of $K$.
The S/N can then be calculated dividing (\ref{eq:SMAX}) by (\ref{eq:NRMS2}).

\section{Measurement results}

\begin{figure}[t]
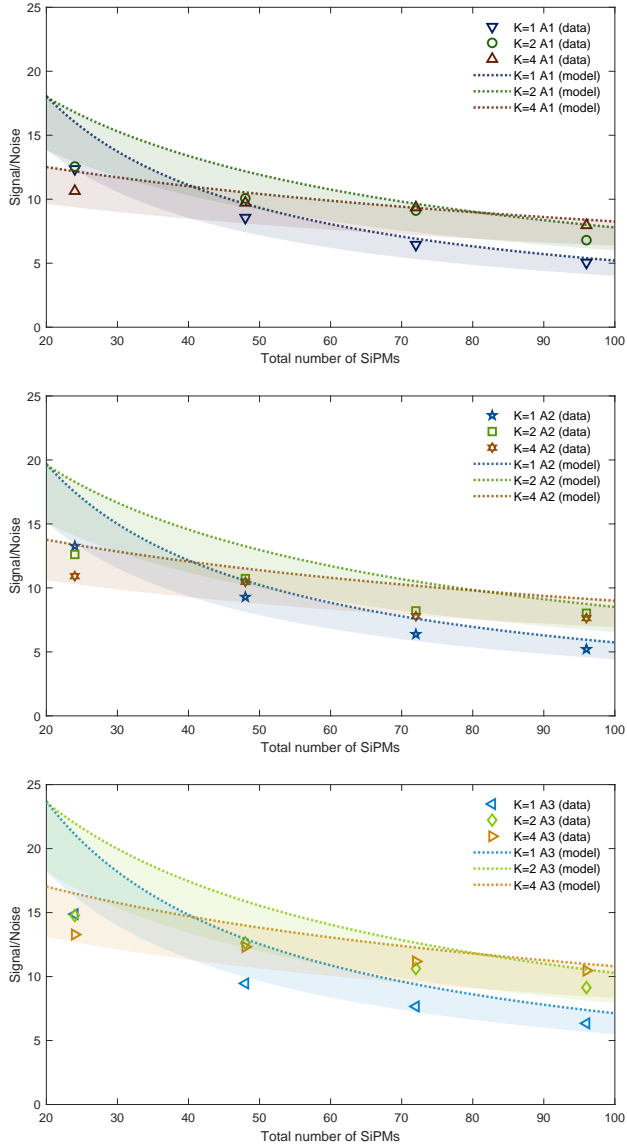

\centering
\includegraphics[width=\linewidth]{SNvsK\_A1.pdf}
\centering
\includegraphics[width=\linewidth]{SNvsK\_A2.pdf}
\centering
\includegraphics[width=\linewidth]{SNvsK\_A3.pdf}
\caption{S/N versus total number of SiPMs in different ganging configurations (K~$=$~1, 2, 4) with the three amplifier variants A1. A2 and A3. The dotted lines represent the calculated values, obtained by dividing (\ref{eq:SMAX}) by (\ref{eq:NRMS2}), with a +30\% tolerance on noise represented by the bands.}
\label{fig:SNvsK}
\end{figure}

Fig. \ref{fig:SNvsK} shows the S/N versus the total number of SiPMs with the three amplifier variants.
The points at 24, 48, 72 and 96 are measured data.
In general a good S/N is measured in all ganging configurations.
The S/N is still decent, equal to 5, even with A1 reading out 96 SiPMs (35~cm$^2$) in parallel, or $K=1$.
This happens because A1 already has a very low series noise.
For the same ganging configuration (same $N$ and $K$), amplifiers A2 and A3 give in general a better S/N than A1, as can be expected.
For all three amplifiers, the S/N with higher $K$ is generally better than that at $K=1$, except for the case at $N=24$ SiPMs. This is due to the presence of the 10~nF input capacitor, which increases the noise gain when the source impedance of the SiPM array is high, as discussed in section \ref{subsec:10nFcap}.
With $N=48$ and more SiPMs the effect of said capacitor becomes less noticeable, and the S/N is consistently better with $K=2$ and $K=4$ than with $K=1$.

The plots in Fig. \ref{fig:SNvsK} also show dotted lines corresponding to the S/N predicted by the model, obtained by dividing (\ref{eq:SMAX}) by (\ref{eq:NRMS2}).
The bands represent a one-sided +30\% tolerance on the estimated noise.
This accounts for the approximations outlined during the calculations, any uncertainty on the SiPM and amplifier parameters, possible effects of environmental disturbances, as well as on the parameters of the gaussian fits on the photoelectron peaks.
The data points are in a good agreement with predictions, especially for higher number of SiPMs.
The agreement is less satisfactory for the measurements at $N=24$, especially with A3, where it amounts to about 50\%.
The reason was not understood.
Since this is also the measurement that predicts the best S/N, it can be expected to be the most sensitive to other unidentified noise sources, including electromagnetic interference from the environment.
Despite the discrepancy, the model consistently predicts the relative ordering of the S/N with different values of $K$.
The agreement between measured and expected values becomes much better as the number of SiPMs is increased.

\begin{figure}[t]
\centering
\includegraphics[width=\linewidth]{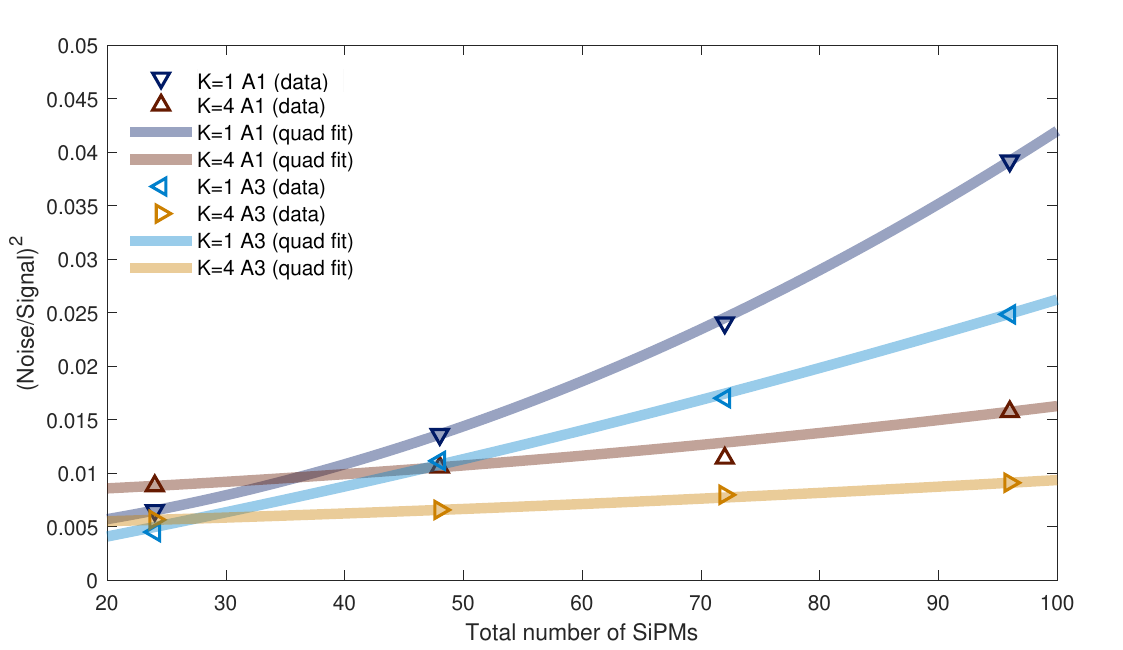}
\caption{(N/S)$^2$ for the same data of Fig. \ref{fig:SNvsK}. The curves are quadratic fits highlighting the behaviour of the measured data versus the total number of SiPMs.}
\label{fig:NS2vsK}
\end{figure}

Fig. \ref{fig:NS2vsK} shows the square of noise over signal, (N/S)$^2$, for the same data as in Fig. \ref{fig:SNvsK}, omitting the intermediate cases $K=2$ and amplifier $A2$.
Showing (N/S)$^2$ makes more evident if the S/N is dominated by the noise of the amplifier, in which case (N/S)$^2$~$\sim N^2$ {\rev (see (\ref{eq:SNsmallK}))}, or by the noise of the quenching resistors, in which case (N/S)$^2$~$\sim N$ {\rev (see (\ref{eq:SNlargeK}))}.
The curves are quadratic fits to the data.
For the same ganging configuration, measurements with an amplifier with lower noise give a consistently better S/N (lower N/S). But changing the ganging configuration is more effective. For $N=96$ the measurement with A1 and $K=4$ gives a better S/N than that with A3 and $K=1$.
Moreover, the (N/S)$^{2}$ with $K=1$ shows a distinct quadratic behaviour with the total number of SiPMs, as was expected from (\ref{eq:SNsmallK}), since $K <\tilde{K}$ for both A1 and A3.
Conversely, measurements with $K=4$ exhibit a linear behaviour of (N/S)$^{2}$ with the total number of SiPMs, as expected from (\ref{eq:SNlargeK}), since now $K >\tilde{K}$, at least in this range.
In other words with $K=4$ and both A1 and A3 the S/N scales as $1/\sqrt{N}$, which confirms the expected behaviour.

\section{Conclusions}
This paper presented a general framework to determine the optimal ganging configuration to read out large arrays of SiPMs with a single amplifier, while preserving the capability to detect single photons.
Given the relevant parameters of SiPM and amplifier, it consists in setting the value of the parameter $K$, indicating the number of SiPMs connected in series through capacitors, to be above $\tilde{K}$ as expressed by (\ref{eq:tildeK}).
Using an amplifier with low voltage noise gives a low $\tilde{K}$, making it possible to reach the optimal regime where the main noise contributor is the quenching resistance of the SiPMs, and the S/N becomes inversely proportional to the square root of the total photosensitive area.

The paper discussed an amplifier design that uses a SiGe transistor as the input device, or two transistors in parallel to further reduce the voltage noise.
It operates at cryogenic temperature with a low power consumption and is particularly well suited for this purpose.
The theory was tested with a set of measurements with different ganging schemes and amplifier variants, observing a good match with the predicted behaviour.


{\rev
\section*{Appendix A}
Integration on a time window of finite length $T$ centered at time $t_0$ (essentially equivalent to a moving average) is represented by convolution with the boxcar function
\begin{equation}
I(t) = \theta \left[t - \left(t_0 - \frac{T}{2}\right) \right] -  \theta \left[t - \left(t_0 + \frac{T}{2}\right) \right]
\end{equation}
where $\theta(t)$ is the Heaviside step function.
This is the impulse response of the filter.
The Fourier transform gives its transfer function:
\begin{equation}
\nonumber \tilde{I}(\omega) = T \frac{\sin \left(\omega T/2 \right)}{\omega T/2} e^{- i \omega t_0}
\end{equation}
The -3 dB cutoff frequency is where
$\nonumber |\tilde{I}(\omega) |^2 / |\tilde{I}(0) |^2 = 1/2$, which occurs for $\omega T/2 \simeq 1.39$ or $f \simeq 0.443/T$.
Integration over a window $T$ can then be approximated in the Laplace domain by a single-pole low pass filter with transfer function
\begin{equation}
T(s) \simeq \frac{\tau_i}{1+s\tau_i}
\end{equation}
where $\tau_i \simeq T/(2 \pi \cdot 0.443) \simeq T/2.78$.

\section*{Appendix B}
Passing from (\ref{eq:Nw}), (\ref{eq:N1f}) and (\ref{eq:Ni}) to the RMS fluctuation (\ref{eq:NRMS}) via Parseval's theorem makes use of the following integrals:
\begin{equation}
\nonumber \int_0^\infty \frac{ (\omega \tau^2 )^2}{(1+\omega^2\tau^2)^2} \frac{d \omega}{2 \pi}= \frac{\tau}{8}
\end{equation}
\begin{equation}
\nonumber \int_0^\infty \frac{ \omega \tau^4}{(1+\omega^2\tau^2)^2} \frac{d \omega}{2 \pi}= \frac{\tau^2}{4 \pi}
\end{equation}
\begin{equation}
\nonumber \int_0^\infty \frac{\tau^2}{(1+\omega^2\tau^2)} \frac{d \omega}{2 \pi}= \frac{\tau}{4}
\end{equation}

}

\ifCLASSOPTIONcaptionsoff
  \newpage
\fi

%




\end{document}